\documentclass[conference]{IEEEtran}
\IEEEoverridecommandlockouts

\usepackage{cite}
\usepackage{amsmath,amssymb,amsfonts}
\usepackage{algorithmic}
\usepackage{graphicx}
\usepackage{textcomp}
\usepackage{xcolor}
\usepackage{url}
\def\BibTeX{{\rm B\kern-.05em{\sc i\kern-.025em b}\kern-.08em
    T\kern-.1667em\lower.7ex\hbox{E}\kern-.125emX}}
\begin{document} 

\title{Technological Progress and Obsolescence: Analyzing the Environmental \& Economic Impacts of MacBook Pro I/O Devices\\
}

\author{\IEEEauthorblockN{Yun-Chieh Cheng}
\IEEEauthorblockA{\textit{College of Engineering} \\
\textit{Ohio State University}\\
Columbus, Ohio \\
cheng.1995@osu.edu}
\and
\IEEEauthorblockN{Yu-Tong Shen}
\IEEEauthorblockA{\textit{College of Engineering} \\
\textit{Ohio State University}\\
Columbus, Ohio \\
shen.1886@osu.edu}
\and
\IEEEauthorblockN{Guanqun Song}
\IEEEauthorblockA{\textit{College of Engineering} \\
\textit{Ohio State University}\\
Columbus, Ohio \\
song.2107@osu.edu}
\and
\IEEEauthorblockN{Ting Zhu}
\IEEEauthorblockA{\textit{College of Engineering} \\
\textit{Ohio State University}\\
Columbus, Ohio \\
zhu.3445@osu.edu}
}

 \maketitle
\begin{abstract}
This study investigates how the new release of MacBook Pro I/O devices affects the obsolescence of related accessories. We also explore how these accessories will impact the environment and the economic consequences. As technology progresses, each new MacBook Pro releases outdated prior accessories, making more electronic waste. This phenomenon makes modern people need to change their traditional consumption patterns. We analyze changes in I/O ports and compatibility between MacBook Pro versions to determine which accessories are obsolete and estimate their environmental impact. Our research focuses on the sustainability of current accessories. We explore alternate methods of reusing, recycling, and disposing of these accessories in order to reduce waste and environmental impact. In addition, we will explore the economic consequences of rapid technological advances that make accessories obsolete too quickly. Thereby assessing the impact of such changes on consumers, manufacturers, and the technology industry. This study aims to respond to the rapid advancement of technology while promoting more sustainable approaches to waste management and product design. As the MacBook Pro I/O unit evolves, certain accessories become obsolete with each subsequent version. The purpose of this study is to identify and quantify the environmental and economic impacts of parts end-of-life. We can detect which accessories have become obsolete and assess the environmental impact by comparing I/O port changes and compatibility across MacBook Pro generations. In response to these environmental images, methods are developed to reuse, recycle, and dispose of obsolete accessories to reduce waste and promote sustainable development. Additionally, we evaluate the economic impact of obsolete equipment on consumers and producers.
\end{abstract}

\section{Introduction}
Many people consider the MacBook Pro to be the highest point of technical innovation. It has experienced remarkable evolution throughout time. Its evolution goes beyond feature and design enhancements. It also reaches the input/output (I/O) device domain, causing a shift in the accessory compatibility environment. New features and connectivity possibilities accompany every new Apple product, eventually making some accessories outdated.

The technology of technological products continues to advance and change, and its accessories also need to be eliminated and evolved over time. The evolution of the MacBook Pro not only improves hardware performance but also changes the accessories market. To ensure their accessories work with the newest MacBook Pro models, consumers should always be aware of how new goods work with previous accessories. Even though there will be significant resource waste and occasional discomfort as a result of this transition, modern technology is progressing, and future convenience is increased.

Rapid advancements in science and technology have led to unparalleled levels of creativity, but they have also influenced the environment. With every new release, the possibilities grow, rendering previously manufactured products and outdated accessories useless and generating garbage that negatively affects the environment and the economy.

In the modern world, environmental awareness is a matter that people take very seriously. Getting rid of old accessories will result in more electronic trash, which will have an impact on the environment and the economy. However, developing inventive products could boost the economy. As a result, while technical advancement and the phase-out of outdated items enhance economic growth, the environment suffers concurrently. As such, the appropriate way to handle these antiquated items has emerged as a topic requiring careful consideration.

This research will examine the I/O port changes in the updated version of the MacBook Pro. Our goal is not only to understand how related I/O devices will evolve but also to take a closer look at the impact of this transition. Which accessories will become obsolete as we progress? How many accessories will be eliminated? How will these impact the environment? How can these old accessories be reused, recycled, and disposed of? 
What's the economic impact?

\section{Related Work}
\subsection{Obsoleted product}
Nowadays, people are beginning to pay attention to the environmental sustainability of products, making Obsolete concepts in product design and management important. Consumers are paying attention to the service life of fishery products and whether they can be recycled and reused. We’ve collected four Obsolete-related research papers to explore the impact of obsolescence and what will happen to product life, waste generation, and resource recycling.

In \cite{sierra-fontalvo_deep_2023}, Sierra-Fontalvo et al. explore the nature of various aspects of obsolescence, including technical, functional, psychological, economic, planned obsolescence, and reduction Manufacturing sources and other types. They identified two main design approaches to reducing obsolescence: long-life product design and extended product life design. However, these two methods lack the formal identification of design attributes related to different scrap types, making the researched methods need further research to find ways to solve them. Thus, they hope that future researchers can develop methods to define design attributes. Moreover, it may help change the definition of obsolescence and enhance methods for measuring and predicting different scenarios.

Similarly, Alzaydi \cite{ALZAYDI2024141239} delves into the topic of obsolescence, product design, and product sustainability. She initiates her exploration of the evolution of obsolescence across time before probing into its manifestation as unintended consequences stemming from intentional design choices as well as adjustments to external variables. The research underscores the need to strike a balance between a product's aesthetic appeal and practicality in an effort to prolong its useful life while also calling attention to the ethical, environmental, and financial ramifications of obsolescence.

In addition, because obsolete accessories are closely related to computers, we also focus on the life cycle of computers and their obsolescence trends. In the research paper  \cite{inproceedings}, Yang et al. provide insights into future trends in the sales and production of obsolete computers, focusing on the US market. The study used logistic models to predict obsolete computer penetration, sales, and production. It highlights the importance of smart policy responses to address growing computer waste and highlights the need for sustainable management practices. The obsolete accessories we studied can also be reused by referring to the computer waste policy.

Furthermore, Kastanaki and Giannis provide dynamic estimates of future discarded notebook computer flows and embedded critical raw materials (CRM) in \cite{KASTANAKI202174} Case study perspective. The study assessed the impact of the COVID-19 pandemic on laptop sales in recent years and the resulting increase in obsolete devices. In addition, it highlights the importance of managing end-of-life laptops due to their high content of CRMs and precious metals, as well as the potential for resource recovery and contribution to the circular economy.

These research papers, taken as a whole, focus on how product design and management might solve obsolescence issues in order to reduce wasteful spending, boost resource efficiency, and support sustainability. We can learn about the relationship between obsolescence and observed accessories, obtain techniques to use when examining the trend of observed accessories, and then generate some relevant solution strategies by investigating different facets of obsolescence and suggesting mitigation and management solutions.

\subsection{Environment}
The management of e-waste is important because it has a significant environmental impact. Moreover, it is closely related to human health and ecosystems. We synthesize perspectives from three relevant studies that examine the environmental impacts of e-waste recycling and management practices.

\cite{LIU2023100028} underline the potential influence of electronic waste recycling on environmental degradation. E-waste contains harmful compounds like dioxin, heavy metals, and persistent organic pollutants. When these pollutants enter the air, soil, or water, they pose a direct hazard to human life and disrupt the ecosystem. Moreover, they emphasize the importance of creating green products. Green products can reduce e-waste output and promote advanced recycling technologies, hence reducing environmental contamination. They advocate for anti-pollution legislation and producer take-back schemes; governments must work with producers and customers to prevent environmental damage from e-waste recollection.

Similarly, in \cite{su15031837}, Ghulam et al. look into the environmental consequences of badly handled e-waste. They highlight the annual increase in improperly disposed of e-waste, as well as the presence of hazardous substances in the e-waste stream. They called for an ever-evolving, technically sound, scientifically based e-waste management policy that curbs environmental pollution, fosters a circular economy approach, and is laced with innovation at every step. Their proposal suggested tight regulations over e-waste streams through expanded producer responsibility coupled with 3R strategies overseen by international monitoring organizations — drawing parallels with Liu et al.'s innovative concept.

Moreover, in \cite{ANSHUPRIYA2018103}, Priya and Hait studied the elemental composition of waste printed circuit boards (PCBs) and their beneficiation process. Examining the physicochemical properties of PCBs in different types of end-of-life electrical and electronic equipment (EEE). This study revealed that waste PCBs contain precious metals and rare earth elements (REEs), and their contents were known. The result also validates Liu et al. that electronic waste will produce organic pollutants. Furthermore, they also highlight the economic issues caused by recycling these materials and the potential for secondary sourcing of critical rare earth elements. This will help the electronic waste recycling industry save resources and contribute to the sustainable development of the environment.

Overall, this study examines environmentally acceptable e-waste disposal alternatives. This new device has the potential to help minimize environmental pollution. Furthermore, it prioritizes resource conservation and public health. Thus, green product design is an important method for environmental performance protection since it reduces the production of dangerous compounds. Furthermore, appropriate management techniques and cutting-edge recycling technology can help mitigate the environmental issues associated with e-waste recycling. It teaches us how to safeguard the environment, which we can subsequently apply to concerns produced by obsolete accessories.

\subsection{E-waste}
Electronic waste (e-waste) is an urgent and important global issue requiring comprehensive recycling and resource recovery strategies. How can we effectively recycle and utilize electronic waste so that the environment is not damaged and these wastes can be used and converted into new reusable items?

In the study \cite{met11081313}, Yken et al. pointed out that setting up efficient recycling procedures was important. They showed that there was only a small percentage of e-waste materials are recycled internationally despite their potential for financial benefit. They emphasize the need for better infrastructure and laws and regulations. However, they also highlight the daunting challenges in e-waste processing and recycling, which should underscore the urgency and importance of the issue for our audience. They advocate for consistent collection programs, methodical removal, and material separation processes to increase recycling efficiency.

Similarly, in \cite{williams2008environmental}, Williams et al. discuss how electronic waste should be managed and processed from the environmental, economic, and social aspects. And they particularly focused on personal computers. They highlighted the globalization of reverse supply chains, the significant environmental problems associated with informal recycling operations, and the potential for harm to occupational health. So they propose policy interventions, such as legislating recycling/recycling systems and regulating toxic ingredients, to find ways to reduce the production of these toxic substances, use laws to bind these suppliers and prevent the impact of some corner-cutting recycling processes.

Heacock et al.'s \cite{doi:10.1289/ehp.1509699} emphasize the importance of global cooperation in managing e-waste to mitigate its adverse effects on human health and the environment. They specifically proposed international treaties such as the Basel Convention, the Bangkok Statement, the StEP Initiative, and the Bali Declaration to emphasize the need for the world to understand and prevent the impact of these electronic wastes, as well as methods and the need for intervention and strategies to prevent the environment from getting worse, with special mention of problems that may cause children's health.

These studies explore various aspects of e-waste management and highlight that addressing the impacts of e-waste requires policy development, improved technology, and the promotion of international cooperation, which must be implemented together to address the problem effectively. These studies show many solutions to environmental issues, drivers, and influences for developing sustainable e-waste management strategies.

\subsection{Economy}
Several academic articles have been written about the problem of handling e-waste and its effects on the environment and economics. The author of \cite{murthy2022review} described the problems that outdated electronic devices bring to the world. It highlights the importance of having strong legal structures and innovative technological solutions to promote sustainable and circular economic models. \cite{forti2020global} quantifies the growing amount of e-waste and highlights our urgent need to strengthen recycling measures and practice a circular economy to reduce adverse environmental impacts. \cite{cucchiella2015recycling} discusses more economic factors. It discusses the feasibility and potential benefits of European recycling initiatives and suggests that significant financial benefits can be achieved through improved e-waste recycling procedures. This economic perspective is related to consumer behavior \cite{islam2021global}, which focuses on the importance of consumers being aware of and participating in e-waste management. Furthermore, \cite{awasthi2018modelling} points out the correlation between economic growth and e-waste generation. It also provides some arguments for integrating economic strategies with environmental sustainability efforts. Additionally, \cite{awasthi2018modelling} highlights the link between economic growth and the production of e-waste. It also offers several arguments for combining economic strategies with environmental sustainability initiatives. These studies not only emphasize the multifaceted challenges associated with e-waste but also propose approaches to address e-waste management from many aspects.

\section{Design}

We followed the objective below to achieve our goal:
\begin{itemize}
\item Identify I/O devices and accessories affected by MacBook Pro releases.
\item Quantify obsolete accessories.
\item Assess the environmental impact of obsolete accessories.
\item Explore strategies for reusing, recycling, and disposing of old accessories.
\item Evaluate the economic implications of I/O device obsolescence.
\end{itemize}

\subsection{Identify obsoleted I/O devices}

First, we went to Apple’s website to collect the charging and expansion of MacBook Pro in various periods, then organized all the information into tables, and finally counted the three most important time points: (1) 2015, (2) 2016-2020 and (3) 2021-2023

\begin{figure}[h]
    \centering
    \includegraphics[width=1\linewidth]{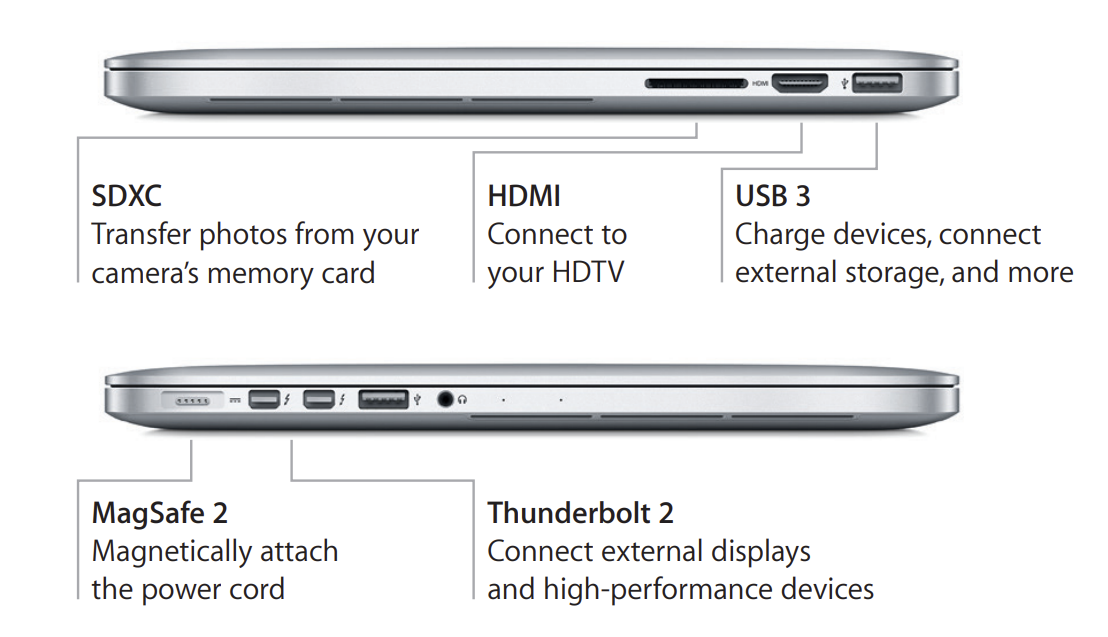}
    \caption{2015 MacBook Pro Charging and Expansion}
    \label{fig:2015_macbook_pro_port}
\end{figure}

Charging and Expansion of 2015 MacBook Pro as shown in Fig~\ref{fig:2015_macbook_pro_port} \cite{2015macbookpro}.
The charging port at that time used MagSafe 2, and it had a complete I/O device port, such as a USB-A port, Headphone port, SDXC card slot, and HDMI port. Based on this comprehensive setup, it appears that no extra adapters are required unless there are particular requirements.

\begin{figure}
    \centering
    \includegraphics[width=1\linewidth]{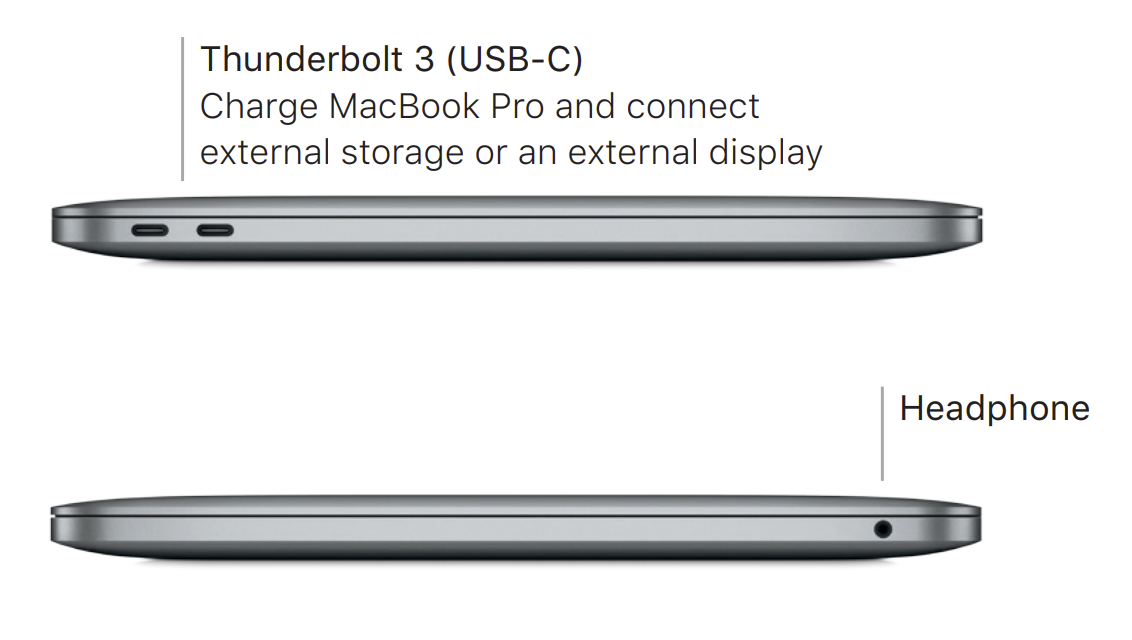}
    \caption{2016 MacBook Pro Charging and Expansion}
    \label{fig:2016_macbook_pro_port}
\end{figure}

Subsequently, between 2016 and 2020 (Fig~\ref{fig:2016_macbook_pro_port})\cite{2016macbookpro}, the MacBook Pro's charging port was replaced with Thunderbolt 3 (USB-C) connectors. Most of the I/O ports have been removed in preference for USB-C slots, which means that we'll need to buy an additional adaptor to use the remaining ports. For instance, you would need to buy an additional "USB-C to SD Card Reader" to read the data if you needed to read the SD card.

\begin{figure}[h]
    \centering
    \includegraphics[width=1\linewidth]{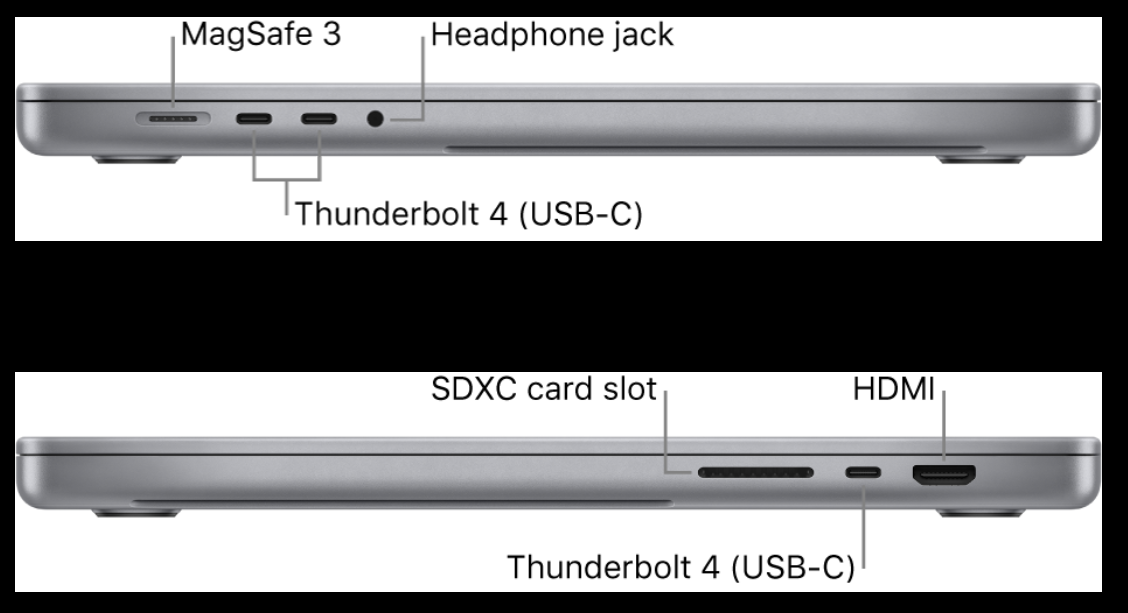}
    \caption{2021 MacBook Pro Charging and Expansion}
    \label{fig:2021_macbook_pro_port}
\end{figure}

The latest MacBook Pro (from 2021 to now) has replaced the charging port of the MagSafe series and upgraded to MagSafe 3. Additionally, the HDMI connector and SDXC card slot have been put back, which can lessen the need for adapters. The most recent I/O port is visible in Fig~\ref{fig:2021_macbook_pro_port}\cite{2021macbookpro}.

\begin{figure}
    \centering
    \includegraphics[width=1\linewidth]{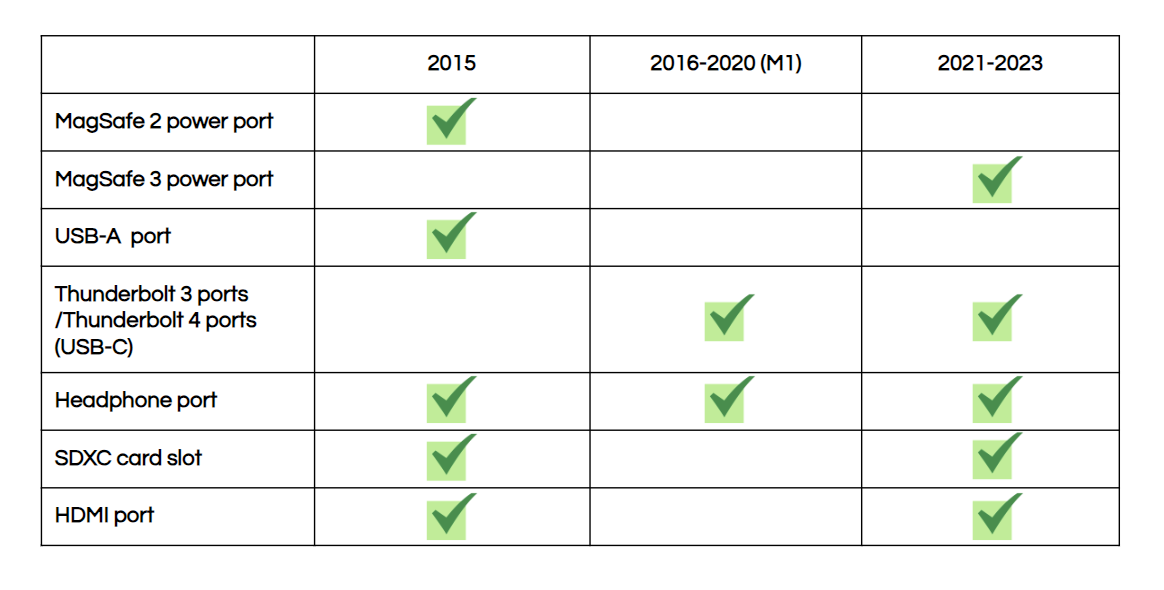}
    \caption{MacBook Pro Port Comparison}
    \label{fig:MacBookProPortComparison}
\end{figure}

After we gathered all the data, we created the comparison chart shown in Fig~\ref{fig:MacBookProPortComparison}. It is visible that the MacBook Pro switches its power port from MagSafe 2 to USB-C and back again. However, MagSafe 2's charger will become outdated because it is incompatible with MagSafe 3.
Furthermore, there hasn't been a USB-A port since 2016, so it cannot use USB-A-related accessories directly.
Since every version has a headphone jack, wired headphones won't become outdated.
Ultimately, there is no need for a separate adapter because the HDMI port and SD card slots were introduced back in 2021.

\begin{figure} [h]
    \centering
    \includegraphics[width=1\linewidth]{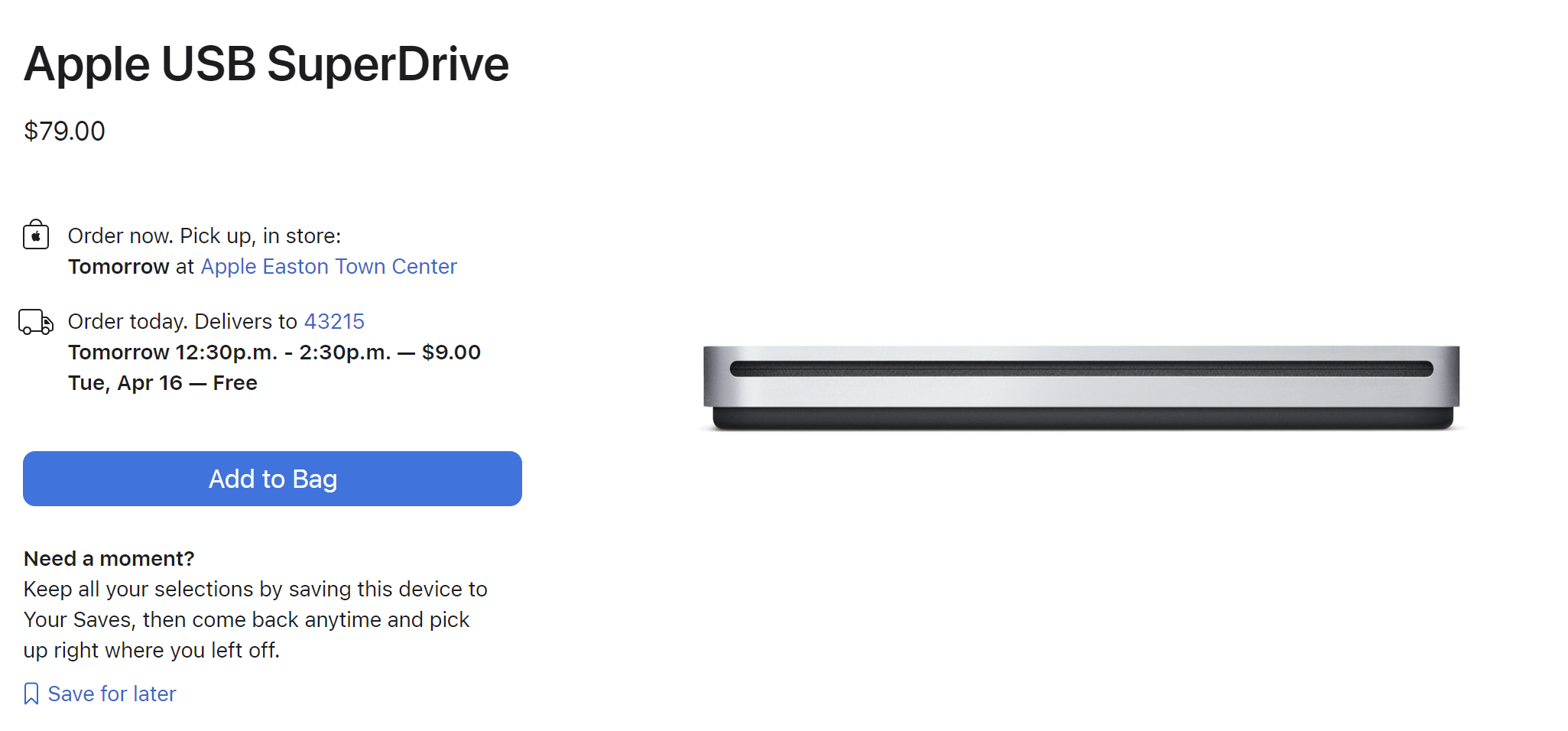}
    \caption{Apple USB SuperDrive}
    \label{fig:SuperDrive}
\end{figure}

\begin{figure} [h]
    \centering
    \includegraphics[width=1\linewidth]{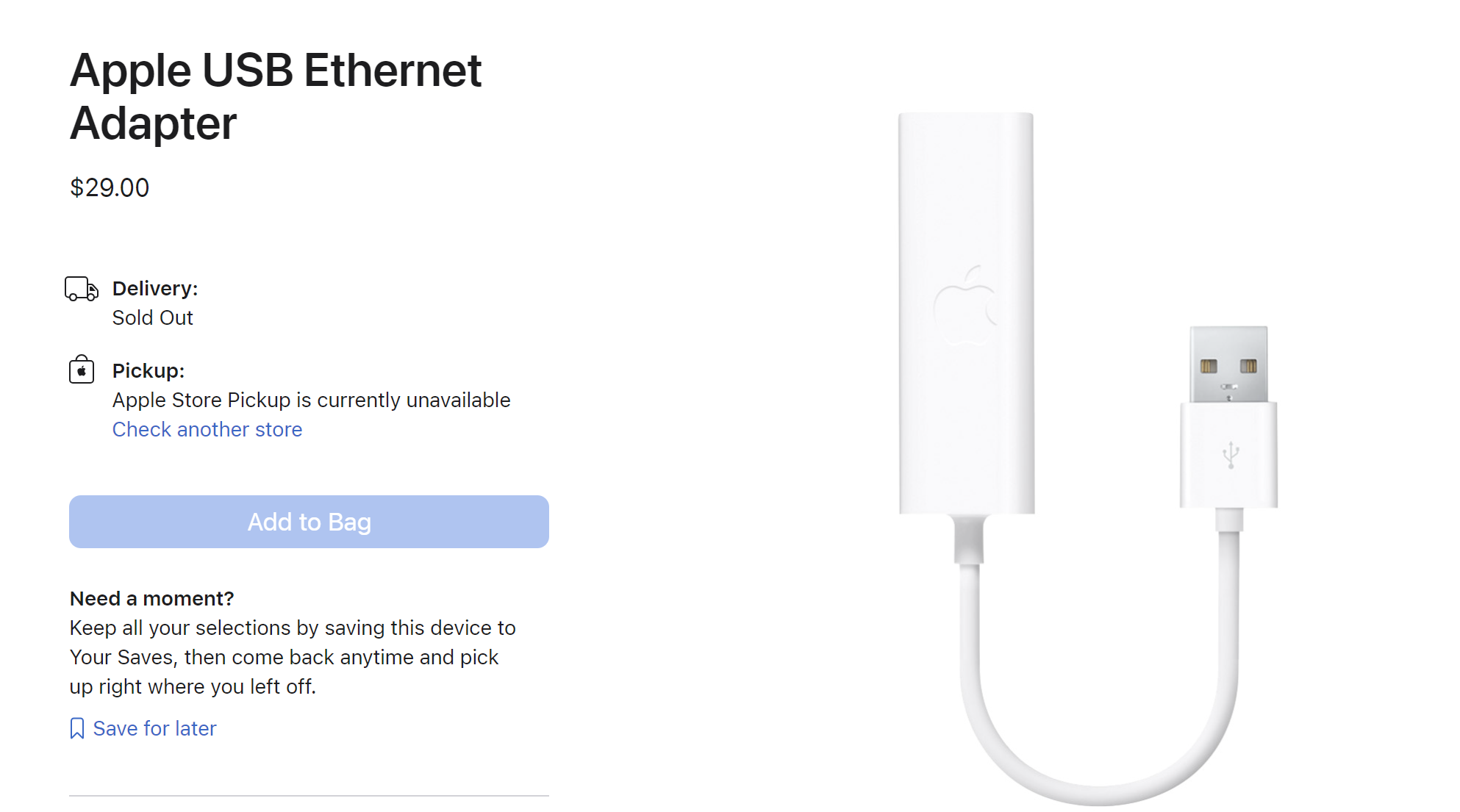}
    \caption{Apple USB Ethernet Adapter}
    \label{fig:ethernet_adapter}
\end{figure}

\begin{figure} [h]
    \centering
    \includegraphics[width=1\linewidth]{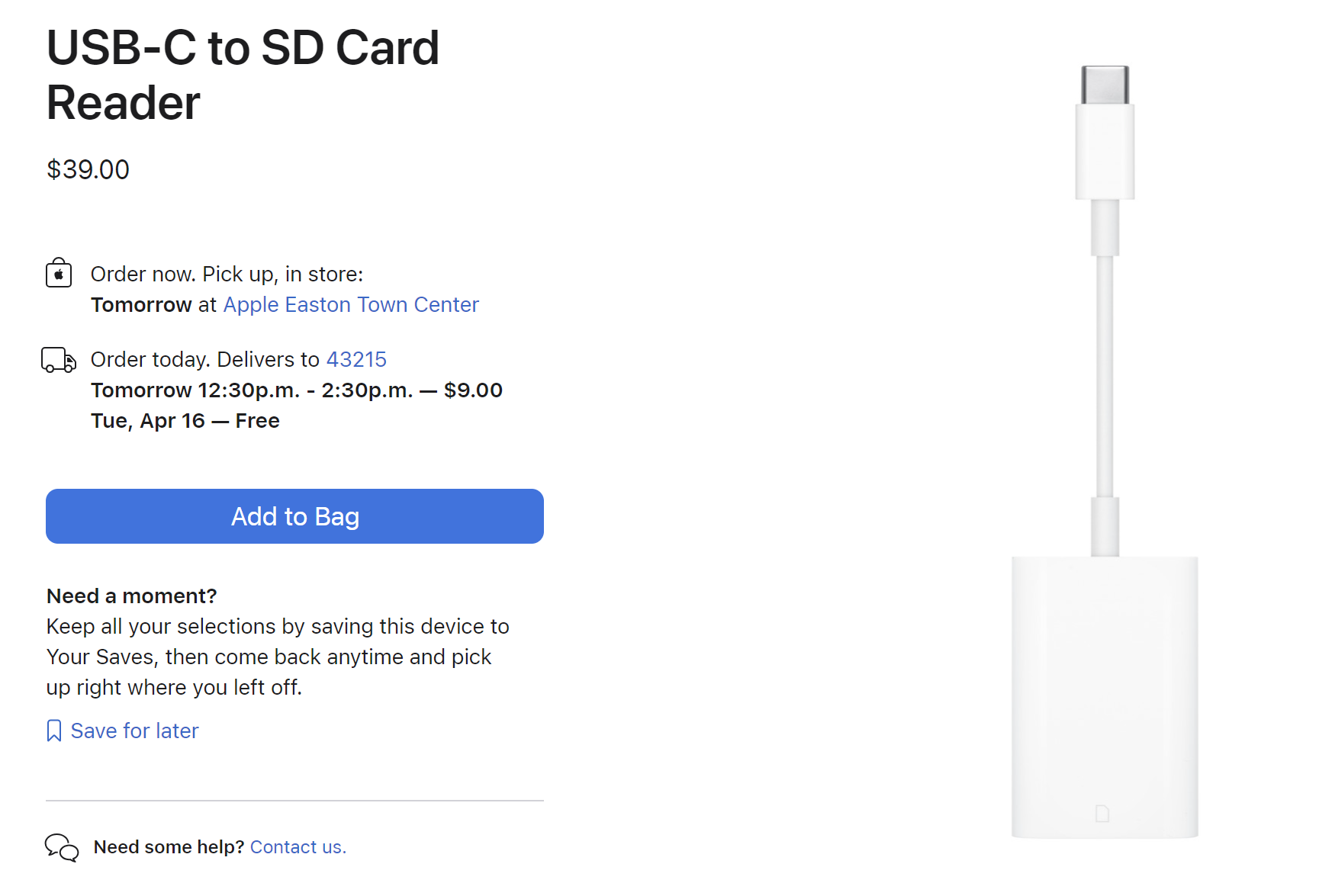}
    \caption{Apple USB-C to SD Card Reader}
    \label{fig:sdcard_reader}
\end{figure}

\begin{figure}
    \centering
    \includegraphics[width=1\linewidth]{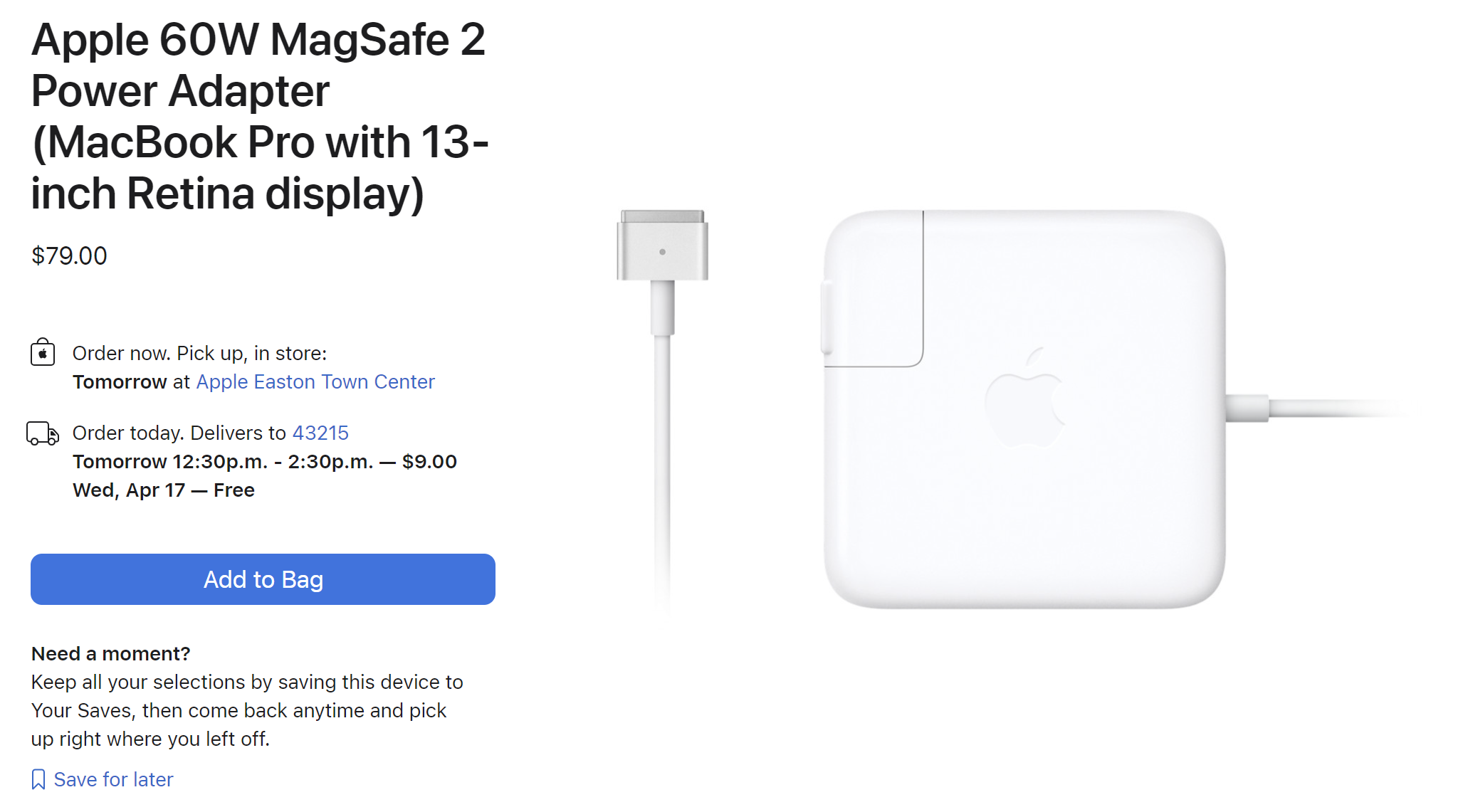}
    \caption{Apple MagSafe 2 Power Adapter}
    \label{fig:magesafe2}
\end{figure}

We visited Apple's official website to look for related accessories that were still available but would eventually become obsolete after comparing every I/O port. These four accessories that we discovered would soon become outdated are:

\begin{itemize}
\item Apple USB SuperDrive  (Fig~\ref{fig:SuperDrive})\cite{AppleUSBSuperDrive}
\item Apple USB Ethernet Adapter (Fig~\ref{fig:ethernet_adapter})\cite{AppleUSBEthernetAdapter}
\item Apple USB-C to SD Card Reader (Fig~\ref{fig:sdcard_reader}) \cite{USB-CtoSDCardReader}
\item MagSafe 2 Power Adapter (Fig~\ref{fig:magesafe2})\cite{AppleMagSafe2}

\end{itemize}

\subsection{Quantify obsoleted I/O devices}

\begin{figure} [h]
    \centering
    \includegraphics[width=0.5\linewidth]{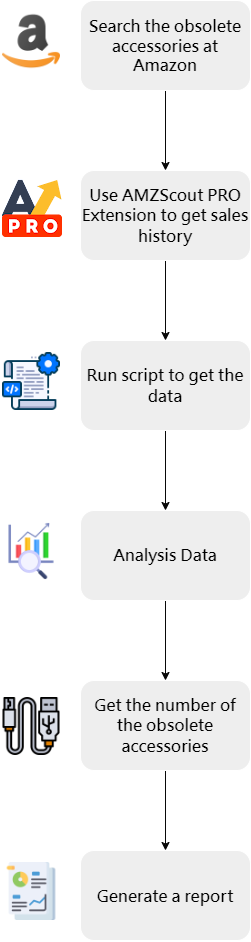}
    \caption{execution process}
    \label{fig:process}
\end{figure}

We require additional information to determine the number of the four obsolete accessories that we have discovered. To determine how many obsolete accessories will be made, we have to determine each product's sales volume. We might consult Fig~\ref{fig:process} as we move forward.

\begin{figure}[h]
    \centering
    \includegraphics[width=1.0\linewidth]{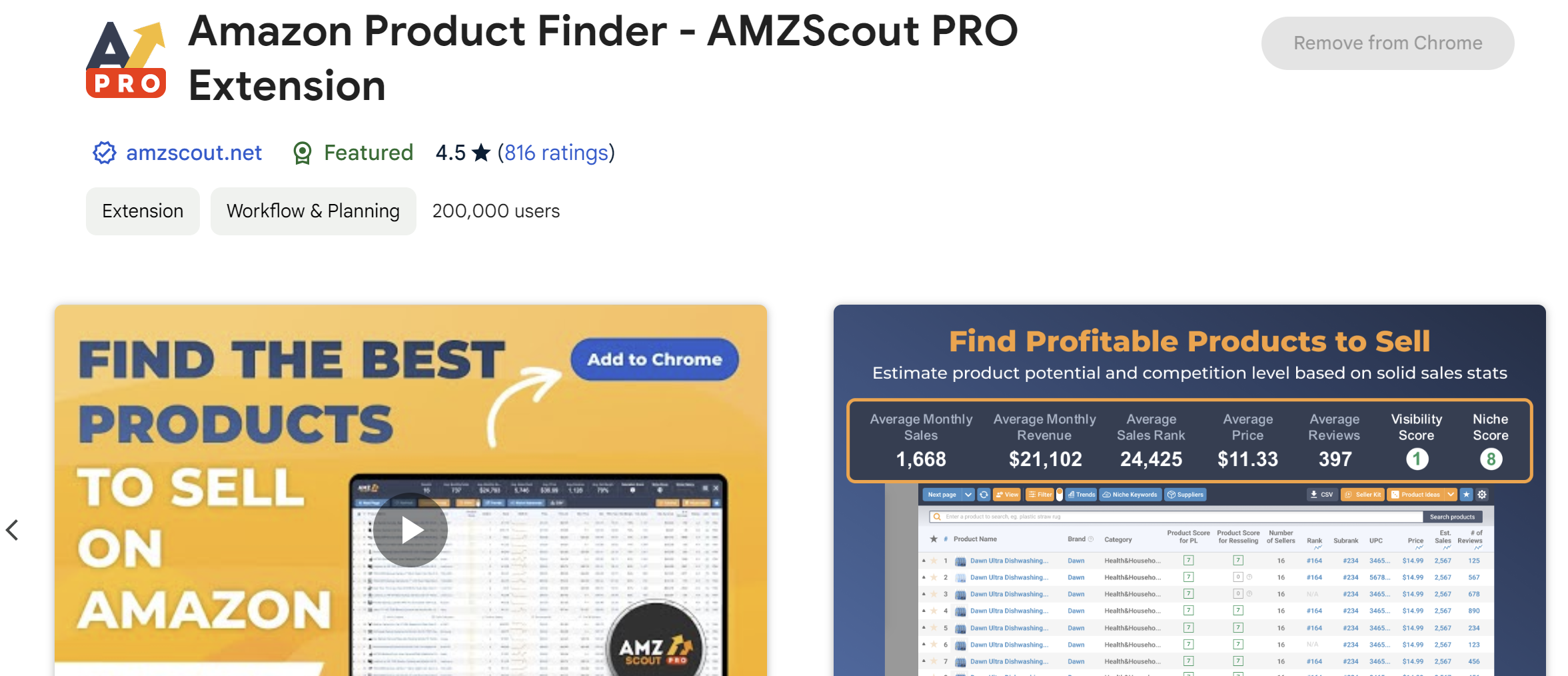}
    \caption{AMZScout PRO}
    \label{fig:amzscout_extension}
\end{figure}

We started by looking for these four I/O devices on Amazon. Since the official Apple website could not provide us with relevant sales data, we opted to gather the data via Amazon. After reaching the product's sales page, we installed a third-party Chrome extension called AMZScout PRO (Fig~\ref{fig:amzscout_extension}). This extension allows us to access the product's sales history directly from Amazon's website. However, the free version of the Extension can only obtain the sales history chart, and we must pay to access the product sales history data CVS file. Therefore, We built an additional script that simulates mouse movements in order to collect comprehensive sales data from the chart.

We used the collected data for analysis once it was finished. To determine when sales were increasing and when they were starting to decline, we first performed some analysis and created comparison charts. Simultaneously, we verified whether the MacBook Pro had been updated and whether any I/O port modifications were made. We may examine the full trend of change from this. Additionally, we will calculate the approximate total sales of these I/O devices and project the number of accessories that will become obsolete when the old MacBook Pro is no longer in use. These accessories will subsequently end up as trash and have a harmful impact on the environment.

\subsection{Research for environmental and economic implications}

\begin{figure}
    \centering
    \includegraphics[width=0.5\linewidth]{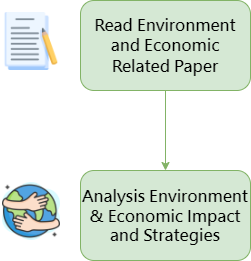}
    \caption{research process}
    \label{fig:research_process}
\end{figure}

Removing these outdated accessories will have specific negative effects on the environment, such as creating a lot of waste, and disposing of this rubbish may cost extra. Aside from the negative effects on the environment, introducing new products will boost the economy because no one will purchase these out-of-date accessories, and recycling or reusing them will be expensive. Thus, we first address the environmental effects of waste, the repurposing and processing of waste materials, and the e-waste problem in our research on the management of outdated accessories. We will talk about the effects of waste and new products on the economy as well as potential future trends. As shown in Fig~\ref{fig:research_process}, first study the relevant research papers. This will facilitate our analysis by revealing some of the environmental damage society knows will cause and some potential impacts. Then, the impacts of these papers will be analyzed. Through these analyses, we will integrate methods and select appropriate strategies to address the impact of premature production of these obsolete parts.

\section{Evaluation}
\subsection{Data Analysis}
In this study, we analyzed sales data for multiple Apple accessories, specifically the Apple MagSafe 2 Power Adapter, Apple USB SuperDrive, and Apple USB-C to SD Card Reader, to evaluate the impact of MacBook Pro hardware updates on the obsolescence of accessories. We focused on sales trends for these accessories. Since some outdated accessories, such as the Apple Ethernet Adapter, are no longer sold in the Apple Store on Amazon and the past data records are incomplete, our analysis primarily focused on the Apple MagSafe 2 Power Adapter, as it has the most comprehensive sales information.

\begin{figure}
    \centering
    \includegraphics[width=1\linewidth]{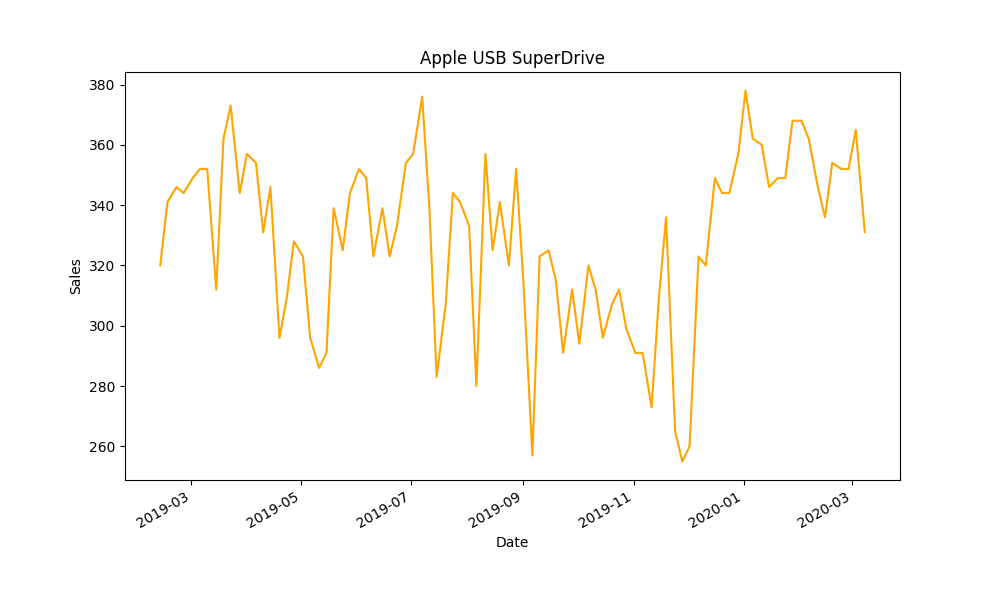}
    \caption{Apple USB SuperDrive Sales}
    \label{fig:Apple_USB_SuperDrive}
\end{figure}

\begin{figure}
    \centering
    \includegraphics[width=1\linewidth]{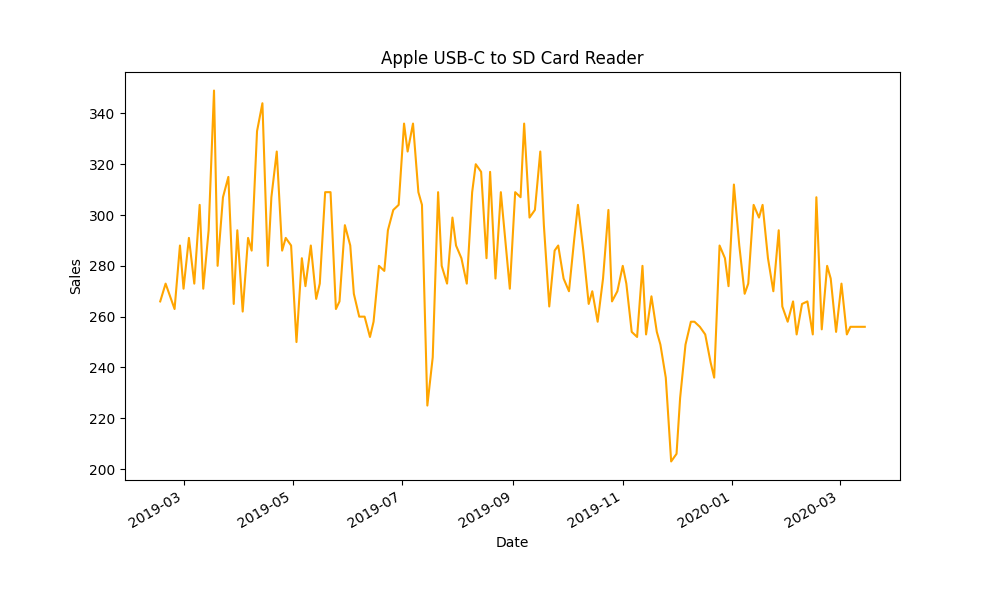}
    \caption{Apple USB-C to SD Card Reader Sales}
    \label{fig:Apple_USB-C_to_SD_Card_Reader}
\end{figure}

We used Python to generate a line chart to visualize the sales trends of these accessories over time. (The sales history of Apple USB SuperDrive is shown in Fig~\ref{fig:Apple_USB_SuperDrive} and Apple USB-C to SD Card Reader is shown in Fig~\ref{fig:Apple_USB-C_to_SD_Card_Reader}.) Data for the Apple MagSafe 2 Power Adapter supply provides a clear timeline of sales fluctuations that can be compared to Apple's product release dates.

\begin{figure}
    \centering
    \includegraphics[width=1\linewidth]{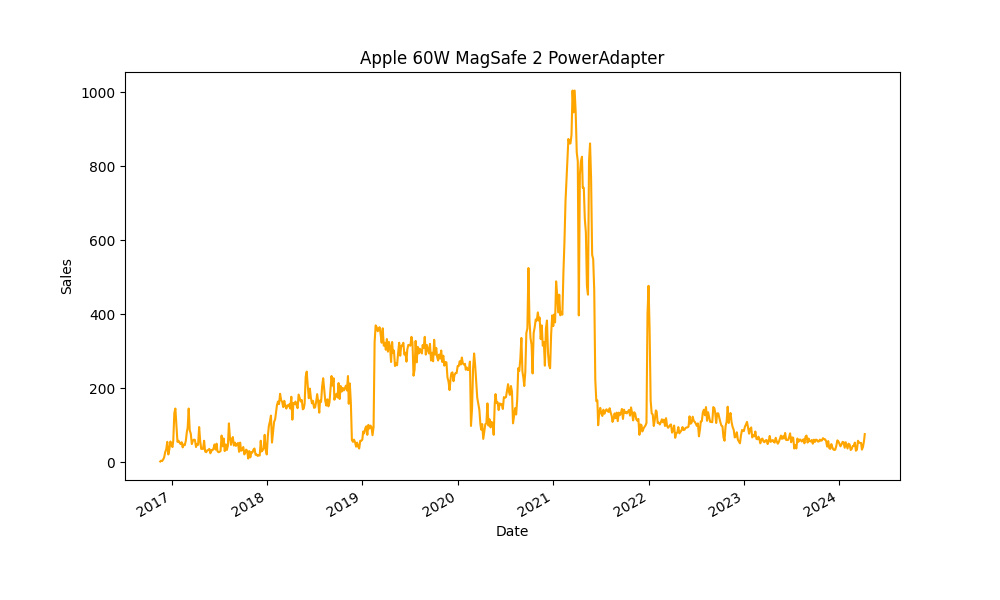}
    \caption{Apple 60W MagSafe 2 PowerAdapter Sales}
    \label{fig:Apple_60W_MagSafe_2_PowerAdapter}
\end{figure}

\begin{figure}
    \centering
    \includegraphics[width=1\linewidth]{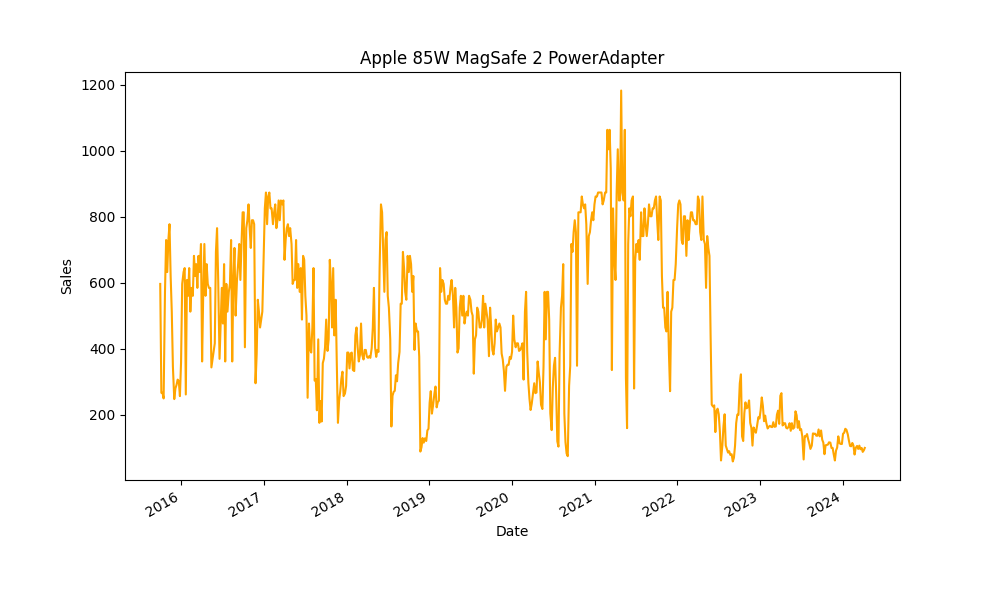}
    \caption{Apple 85W MagSafe 2 PowerAdapter Sales}
    \label{fig:Apple_85W_MagSafe_2_PowerAdapter}
\end{figure}

\begin{figure}
    \centering
    \includegraphics[width=1\linewidth]{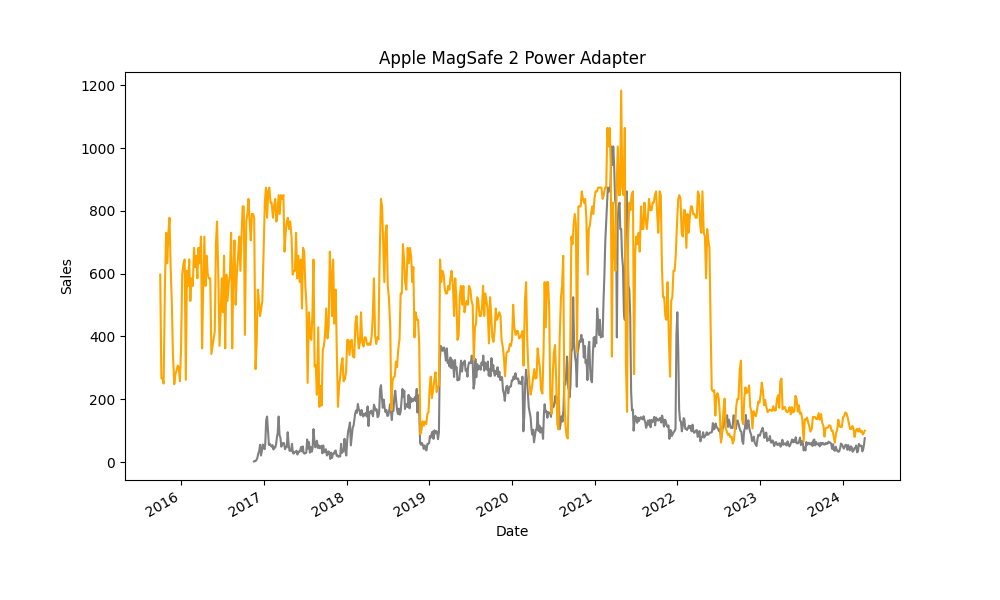}
    \caption{Apple MagSafe 2 Power Adapter Sales}
    \label{fig:Apple_MagSafe_2_Power_Adapter_2lines}
\end{figure}

\begin{figure}
    \centering
    \includegraphics[width=1\linewidth]{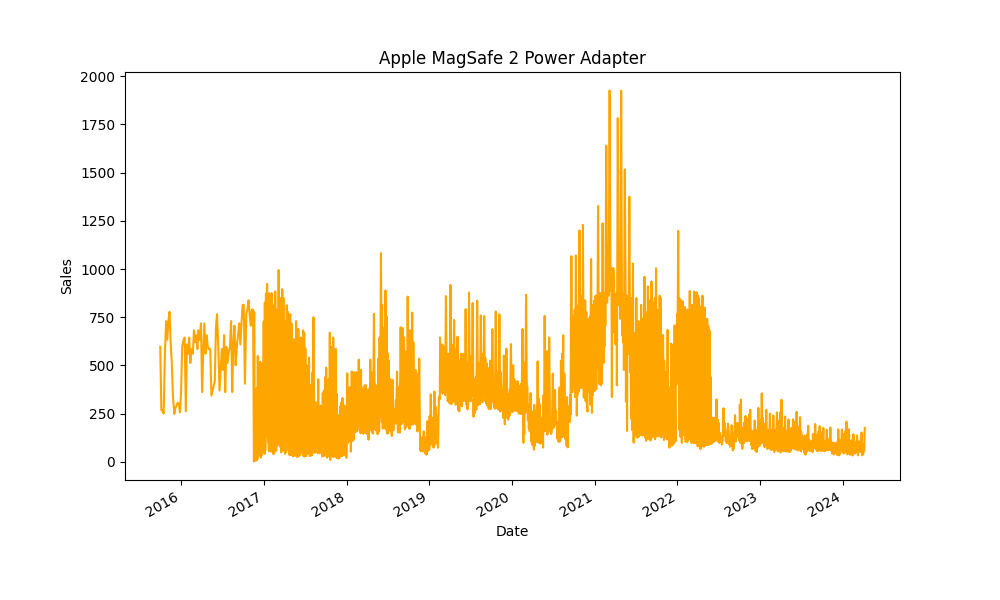}
    \caption{Apple MagSafe 2 Power Adapter Combination Sales}
    \label{fig:Apple_MagSafe_2_Power_Adapter_comb}
\end{figure}

We had data on two wattages of the Apple MagSafe 2 Power Adapter, which we used to generate the line chart. (Fig~\ref{fig:Apple_60W_MagSafe_2_PowerAdapter} and Fig~\ref{fig:Apple_85W_MagSafe_2_PowerAdapter}) Furthermore, the different wattages of MagSafe 2 Power Adapter are the same type as those of ours, so we combined them into one chart. (Fig~\ref{fig:Apple_MagSafe_2_Power_Adapter_2lines} and Fig~\ref{fig:Apple_MagSafe_2_Power_Adapter_comb})

The line chart for the Apple MagSafe 2 power supply shows clear sales peaks and troughs. These fluctuations are closely related to the release of new MacBook Pro models, especially models with significant changes to I/O ports and charging options. For example, immediately after it was announced that an upcoming model would not include a MagSafe 2 port, there was a statistically significant increase in sales. It suggests that people were rushing to buy the adapter before it became obsolete, with consumers buying the adapter either as a replacement or For continued compatibility with older devices.

The analysis shows that Apple's strategic decisions on port configuration can have a direct and measurable impact on accessory sales. This trend highlights the economic impact of outdated accessories. Consumers demand for compatibility or replacement parts may temporarily boost sales of the retirement of older technology. However, as the product eventually becomes irrelevant in newer hardware environments, these peaks are often followed by sharp declines.

For TABLE~\ref{tab:sales_data}, we calculated the sales of all obsolete parts so that we can know how many I/O devices will become trash and no longer needed. The numbers in the table are not exact. The correct number will be larger than what we get. However, we can still find that more than 468,328 accessories will become obsolete, which means that after this time, these accessories will become obsolete and become environmental pollution.

\begin{table}
\centering
\caption{Sales Data for Apple Accessories on Amazon}
\renewcommand{\arraystretch}{1.5} 
\begin{tabular}{|c|c|}
\hline
\textbf{Product} & \textbf{Sales} \\ \hline
Apple 60W MagSafe 2 Power Adapt & 116,591 \\ \hline
Apple 85W MagSafe 2 Power Adapt & 281,241 \\ \hline
Apple USB SuperDrive & 29,621 \\ \hline
Apple USB-C to SD Card Reader & 40,875 \\ \hline
\textbf{SUM} & 468,328 \\ \hline
\end{tabular}
\label{tab:sales_data}
\end{table}

\begin{figure}
    \centering
    \includegraphics[width=1\linewidth]{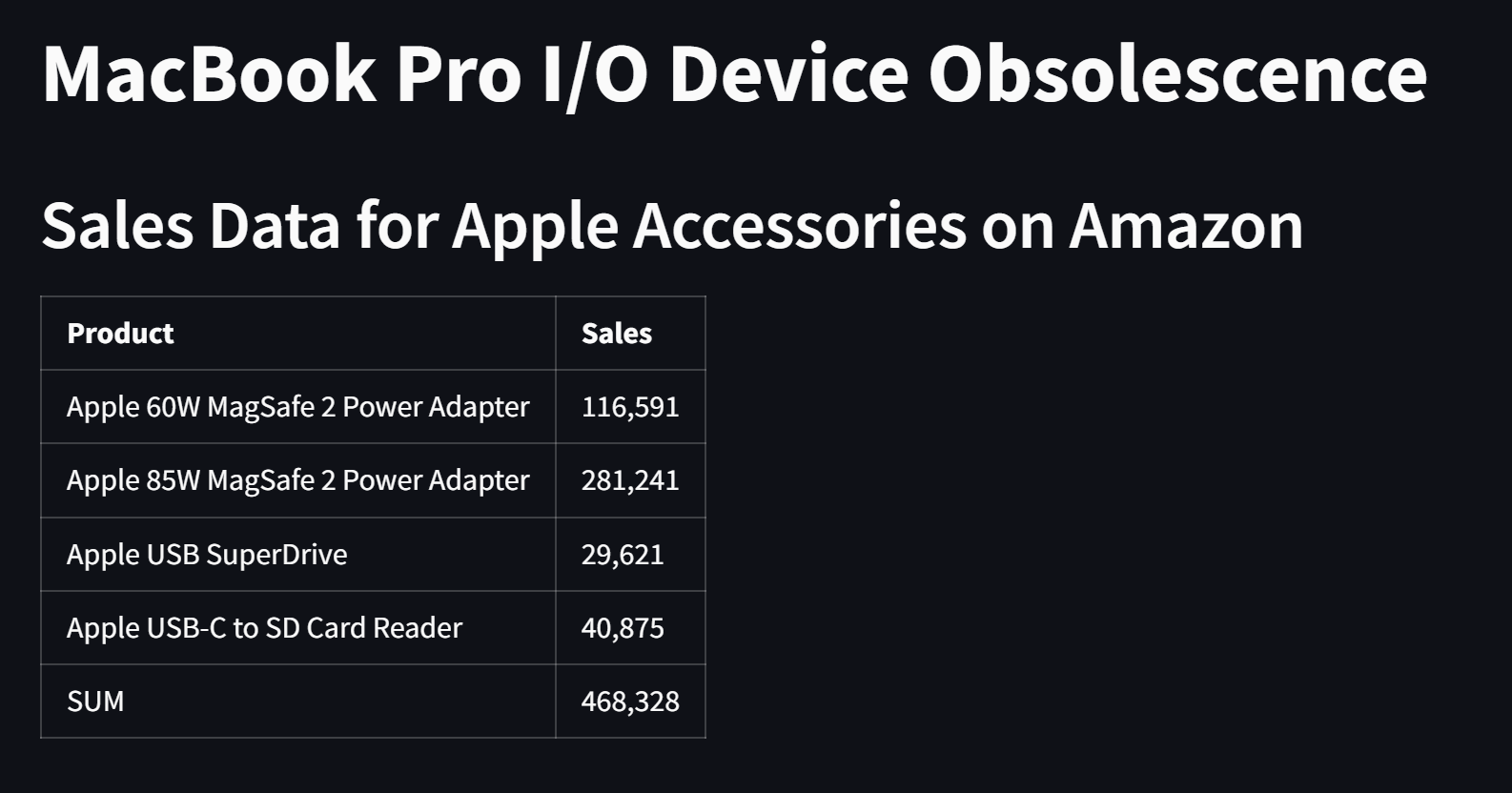}
    \caption{Result on Website -1}
    \label{fig:webpage_table}
\end{figure}

\begin{figure}
    \centering
    \includegraphics[width=1\linewidth]{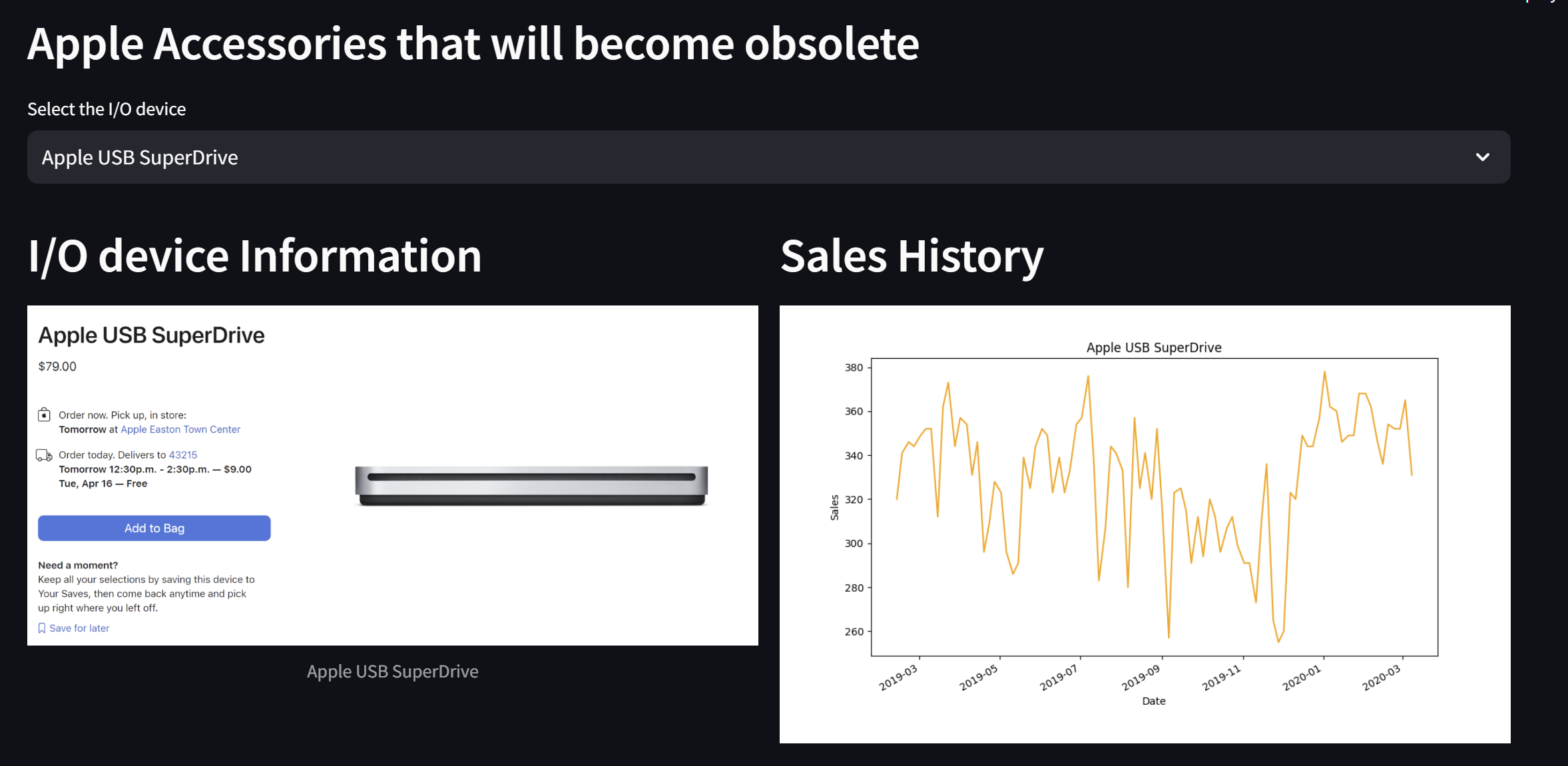}
    \caption{Result on Website -2}
    \label{fig:webpage_chart}
\end{figure}

After generating the chart, we put the chart and the product information on the website, which is more convenient for people to read and analysis with the information. First, we will look at the table showing the sales of these outdated accessories.(Fig~\ref{fig:webpage_table}) Second, we need to select which I/O device we want to research. Then, the website will display the information on the Apple official website about the product on the left and its sales history on the right.(Fig~\ref{fig:webpage_chart}) There will be more than one chat for the Apple MagSafe 2 Power Adapter because we put the chart of two wattages of the Apple MagSafe 2 Power Adapter together and combined the chat.

\subsection{Environmental Impact}
\subsubsection{Increase in Electronic Waste Generation}
Obsolete MacBook Pro I/O devices have led to an increase in electronic waste, posing major environmental risks. Previous I/O devices become obsolete as new MacBook Pro versions or models are released. They must be discarded and replaced \cite{JAIN202334}. As a result, this continuous cycle generates a large amount of electronic trash. So, we must be carefully controlled and disposed of.

\subsubsection{Resource Depletion and Energy Consumption}
The manufacturing process of new I/O devices to replace obsolete ones requires significant resources and energy \cite{ANSHUPRIYA2018103}. As Priya and Hait emphasize, this process can lead to resource depletion and environmental damage. Extracting raw materials and energy-intensive manufacturing processes further strain our natural resources and exacerbate environmental problems.

\subsubsection{Landfill Contamination}
As discussed in the research \cite{williams2008environmental}, abandoned outmoded I/O devices frequently end up in landfills. These landfills have the potential to pollute the ecosystem by allowing dangerous compounds to leak into the groundwater and soil. This contamination threatens the environment and human health, necessitating proper e-waste management techniques.

\subsubsection{Air and Water Pollution}
If Improper handling and recycling of e-waste, including obsoleted I/O devices, can lead to air and water pollution \cite{JAIN202334}. The incineration of e-waste releases harmful pollutants into the air, and it will cause air pollution. In addition, leachate from landfills contains e-waste components. They can contaminate nearby water sources and harm ecosystems and human health.

\subsubsection{Loss of Biodiversity}
As discussed in the research paper by Williams et al., environmental pollution caused by electronic waste disposal may harm biodiversity\cite{williams2008environmental}. Releasing toxic substances into the ecosystem will destroy the fragile ecological balance. Moreover, it will threaten the survival of various species. Protecting biodiversity is important to maintaining healthy ecosystems and ensuring the survival of all life on Earth.

\subsubsection{Climate Change Impact}
According to Priya and Hait's study, the manufacture and disposal of electronic equipment, including I/O devices, emits greenhouse gases \cite{ANSHUPRIYA2018103}. These emissions contribute to climate change and cause various environmental and socioeconomic consequences. We must address these emissions through sustainable activities. To prevent climate change and protect the world for future generations.

All in all, the obsolescence of MacBook Pro I/O devices will profoundly impact the environment, and we need to pay attention to sustainable practices and implement effective e-waste management strategies.

\subsection{Strategies}
\subsubsection{Promotion of Extended Producer Responsibility (EPR) Programs}
Implementing an EPR program can encourage manufacturers to take responsibility for their products' whole life cycle, including disposal and recycling \cite{su15031837}. The EPR program reduces electronic waste by requiring manufacturers to account for the environmental impact of their products and encouraging the development of more durable and recyclable equipment.

\subsubsection{Development of Circular Economy Practices}
Adopting circular economy principles helps to reduce resource consumption and waste generation related to electronic device end-of-life \cite{JAIN202334}. A circular economy approach can reduce environmental effects by encouraging product reuse, repair, and recycling. This extends the service life of I/O devices and facilitates the recycling of valuable components.

\subsubsection{Advancement of Green Product Design}
By encouraging the design of green products, greener I/O devices can be developed \cite{LIU2023100028}. Products with modular components, recyclable materials, and energy-saving features can reduce the environmental damage caused by electronic devices and promote sustainability.

\subsubsection{Expansion of E-waste Recycling Infrastructure}
Investing in expanding e-waste recycling infrastructure can improve the proper management and disposal of discarded I/O devices \cite{met11081313}. By increasing the availability of recycling facilities and public awareness of e-waste recycling options, more devices can be diverted from landfills and recycled. And then it can reduce environmental pollution.

\subsubsection{Enforcement of Strict Environmental Regulations}
Strengthening environmental regulations related to electronic waste management can help mitigate the adverse environmental impacts of obsolescence \cite{williams2008environmental}. Governments can ensure compliance with environmentally sustainable practices by enforcing regulations on e-waste disposal, hazardous substance management, and recycling practices.

\subsubsection{Encouragement of Consumer Education and Behavior Change}
Educating consumers about the environmental impacts of electronic waste and promoting sustainable consumption habits can foster a culture of responsible consumption \cite{ANSHUPRIYA2018103}. By encouraging consumers to prolong the lifespan of their devices through repair and reuse, the volume of electronic waste generated can be reduced, leading to positive environmental outcomes.

\subsection{Economic Implications}
Recycling and reusing obsolete accessories has multiple economic benefits, including cost reduction, job creation, and expanded market opportunities. The study says the reuse and recycling industry not only provides jobs but also fosters a market for second-hand equipment, which is particularly beneficial to developing countries as they use cheap technology to boost their economies\cite{williams2008environmental}. In addition, adopting flexible policies means that reuse and recycling rates can be increased during economic expansions with increasing consumption, production, or exports\cite{tsiliyannis2007flexible}. The study also highlights that reuse operations tend to be more profitable than recycling, especially in industries such as mobile phones, suggesting that businesses that focus on reuse are likely to be more financially successful\cite{geyer2010economics}. In addition, increased reuse and recycling of electronic devices can lead to significant cost savings and better resource recovery while also avoiding environmental and health hazards associated with unsuitable disposal\cite{prabhu2023disposal}. By promoting products designed to last longer and be easier to repair, this kind of policy can support economic growth. This way, it doesn’t use up resources too fast or produce too much waste. These benefits highlight the importance of reuse and recycling in promoting a circular economy and economic resilience.

\section{Issues we encountered}
\subsection{Data Collection}
Data collection was the most difficult part of this project. Although it is easy to think about how to get the number of product sales, a lot of methods we have thought do not work. 
First, we wanted to get the number of sales through the Apple official website. However, there was no data we could collect from the product detail page. We had research for a long time. We couldn't get the data from API or other information on the website. Thus, we gave up and changed another way to achieve our goal.

Subsequently, we thought that Apple's official financial statements were helpful information. Maybe we could get the data of the accessories sales on the statement. However, they only displayed the sales of MacBooks or iPhones, these main products; we could not get the sales of especially accessories. We could only get the sales for all accessories. It's not the data that we want.

\begin{figure}
    \centering
    \includegraphics[width=1\linewidth]{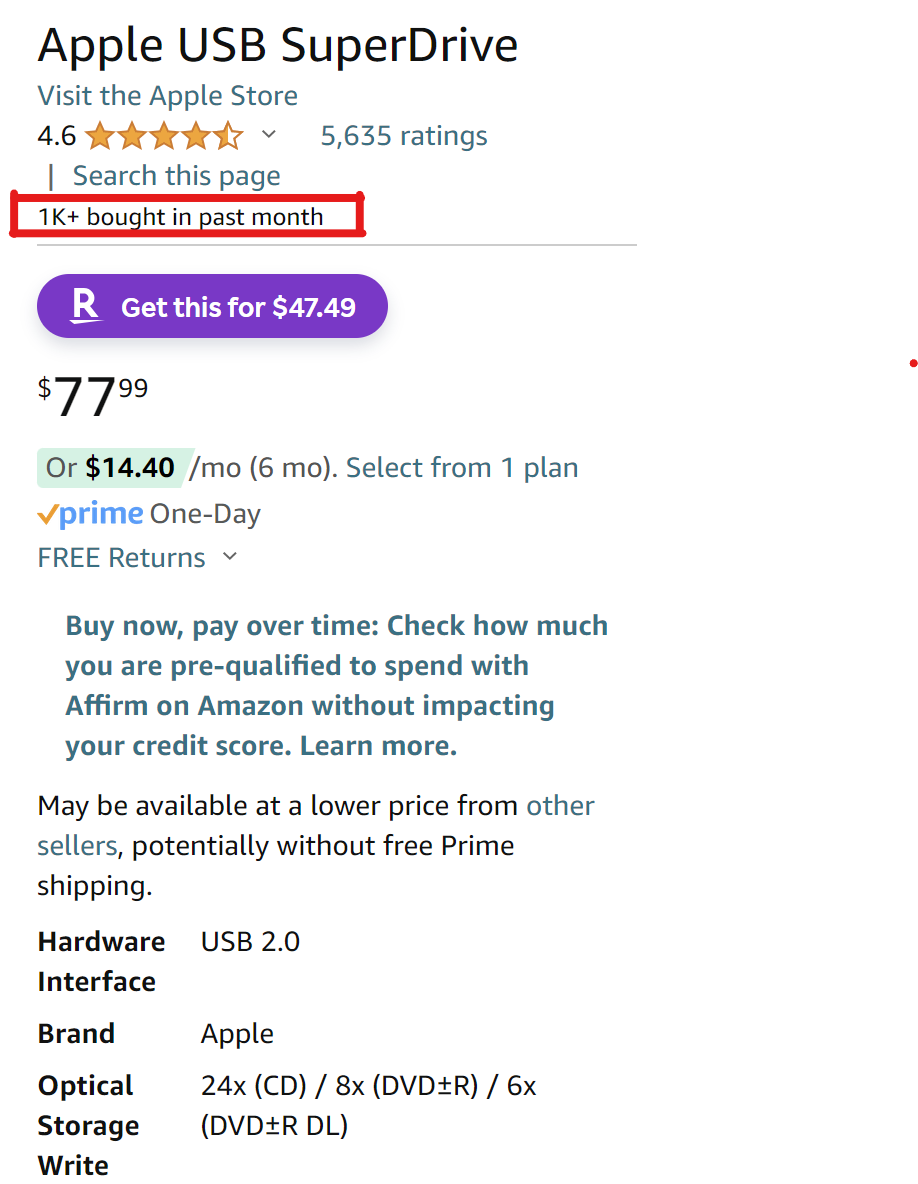}
    \caption{Amazon sales history}
    \label{fig:amazonsales}
\end{figure}

Therefore, we decided to collect the data from Amazon. At first, we wanted to use crawler technology to obtain data, so we first entered the product page we wanted and then used a script to obtain the information, but we found that Amazon would only display the approximate number of sales last month (Fig~\ref{fig:amazonsales}). Not only was it inaccurate, but there was also only one small amount of data. Thus, we started to find helpful information from the API or other information on the website. And there were only product reviews that were the data that we could use. We needed to find or train a model to analyze the product reviews and sales. It's difficult and inaccurate, so it's not a good solution for us.

\begin{figure}
    \centering
    \includegraphics[width=1\linewidth]{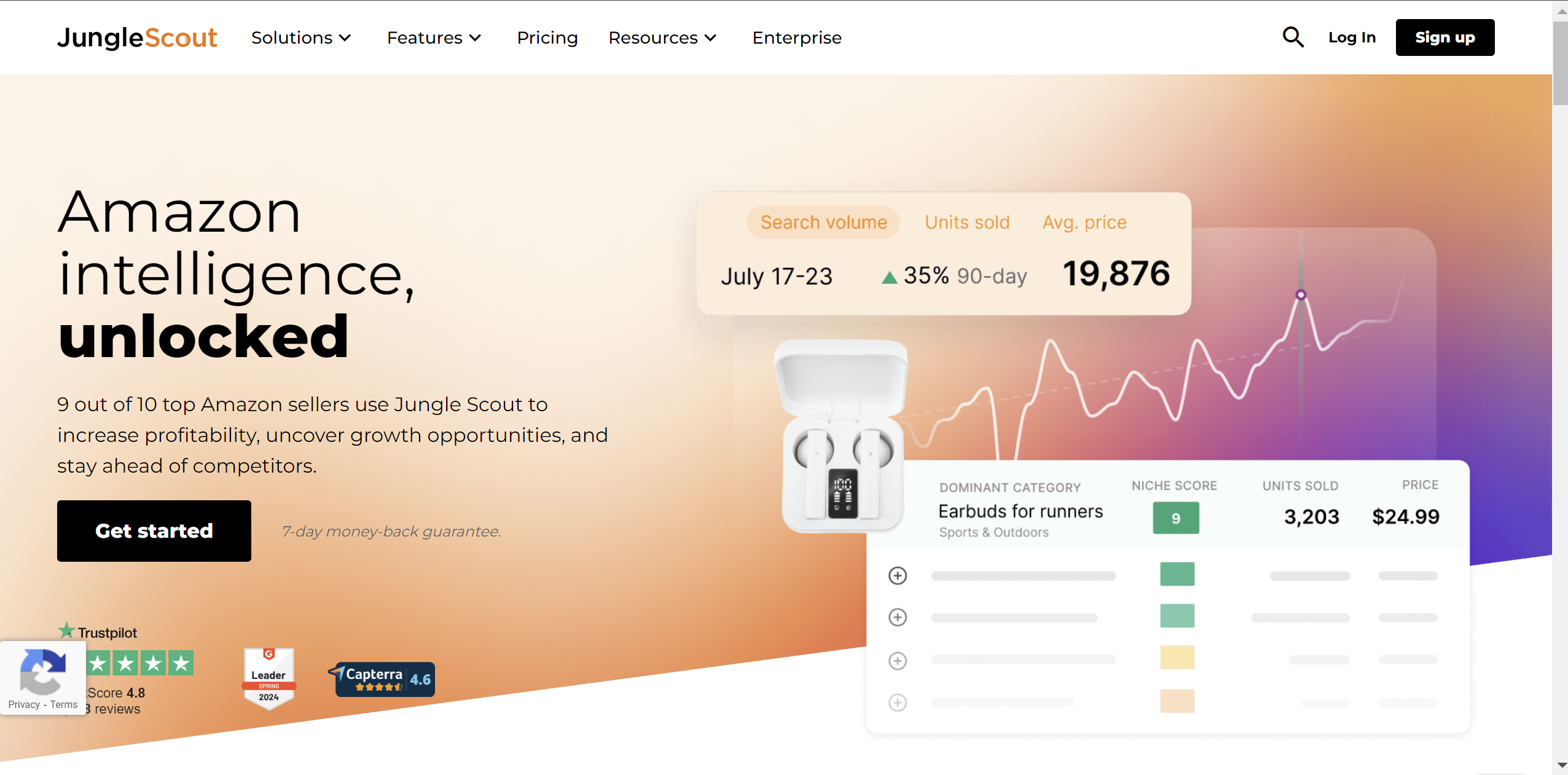}
    \caption{Jungle Scout}
    \label{fig:junglescout}
\end{figure}

\begin{figure}
    \centering
    \includegraphics[width=1\linewidth]{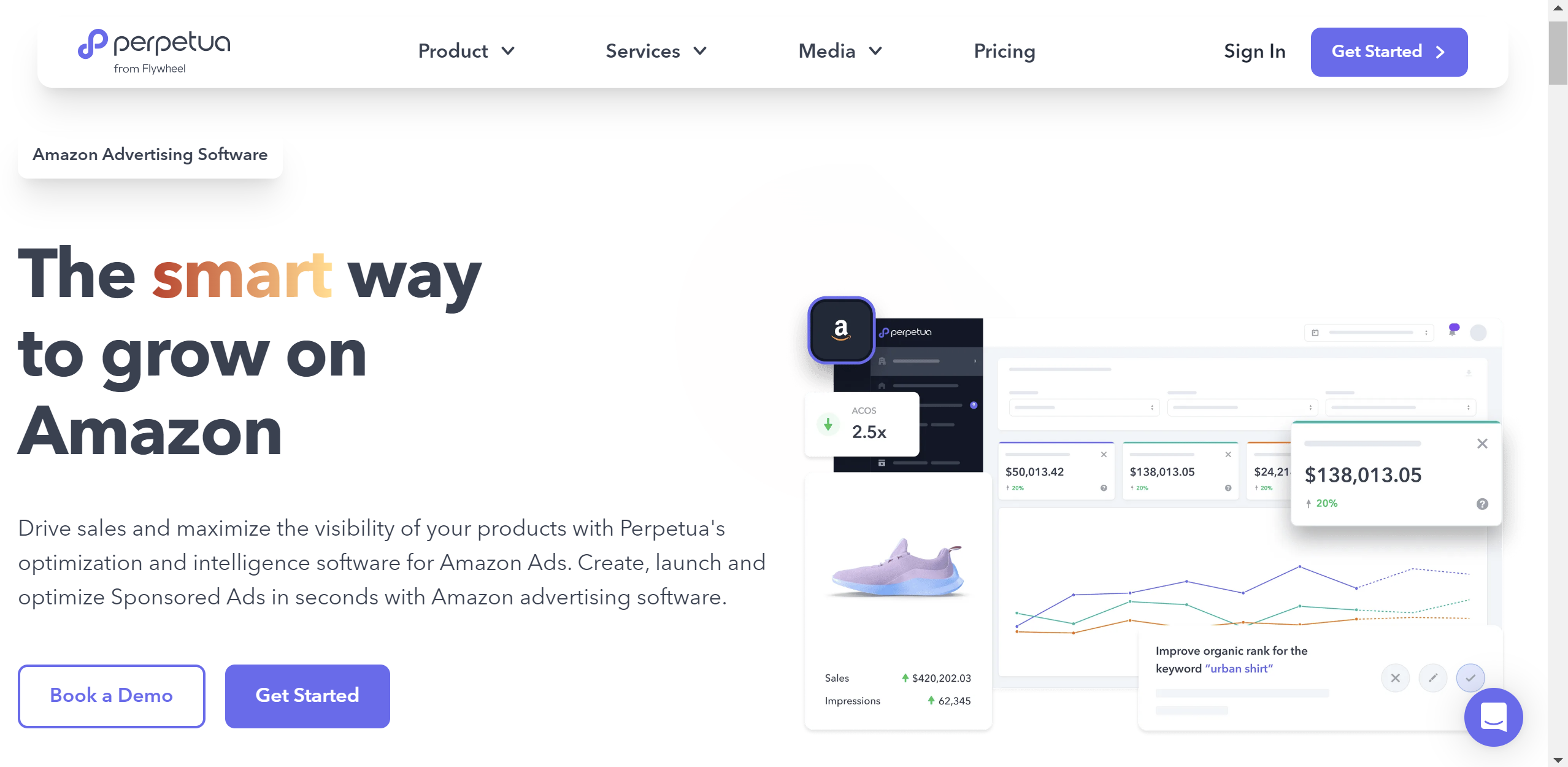}
    \caption{Perpetua}
    \label{fig:perpetua}
\end{figure}

Furthermore, we started looking for some third-party tools to help us collect data. We found some tools, such as Jungle Scout (Fig~\ref{fig:junglescout}) \cite{junglescout} and Perpetua (Fig~\ref{fig:perpetua}) \cite{perpetua}, but some of these tools are for Amazon merchants. We didn't have an account with Amazon merchants, so we could not use them. In addition, these tools require payment to use, and they are expensive that it's not affordable for us. 

\begin{figure}
    \centering
    \includegraphics[width=1\linewidth]{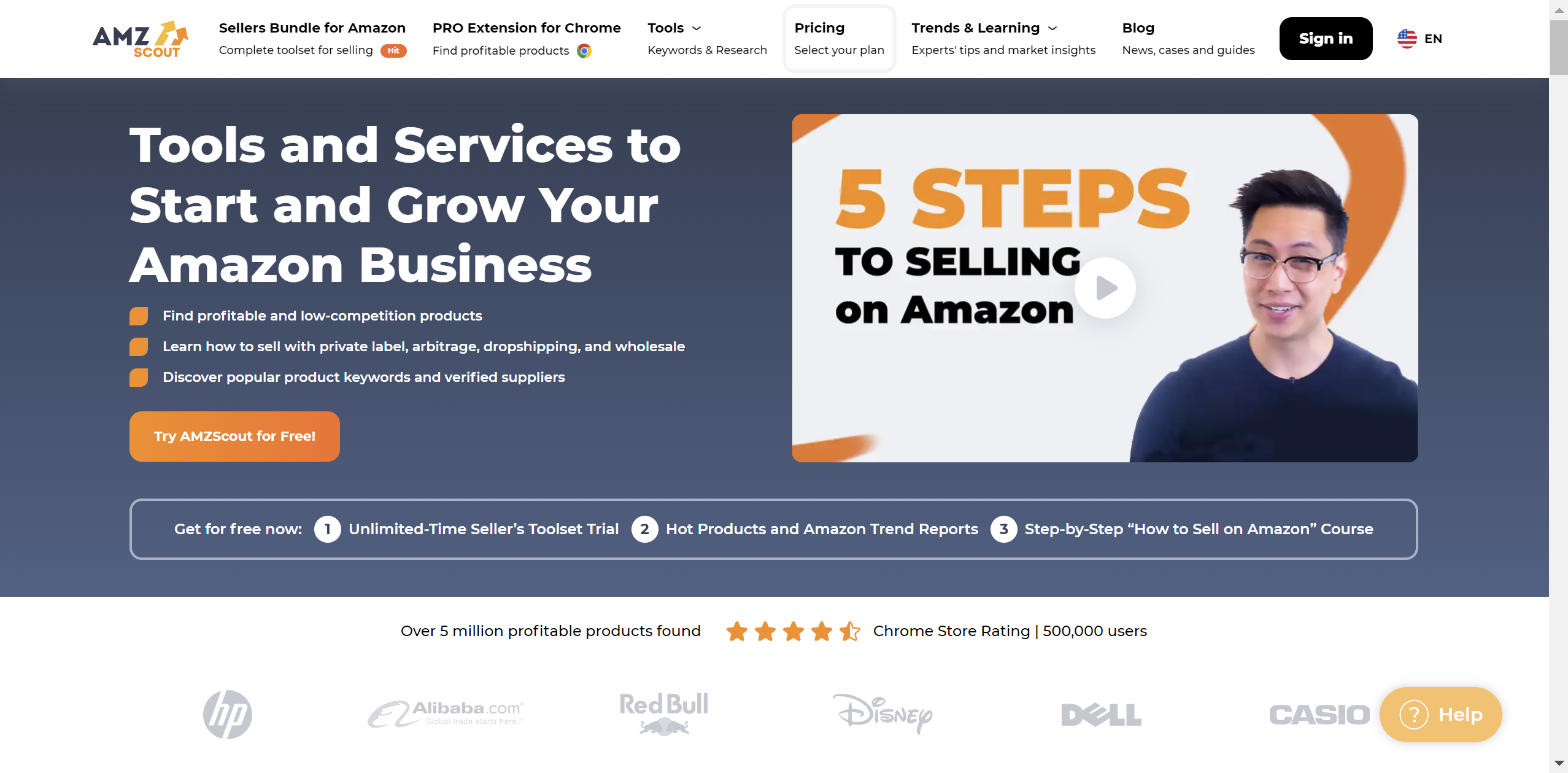}
    \caption{AMZScout}
    \label{fig:amzscout}
\end{figure}

Finally, with our efforts, we found a relatively suitable tool called AMZ Scout (Fig~\ref{fig:amzscout})\cite{amzscout}. When we went to its website, we found that we could search for and see some information about some products, but we could not find the product we wanted on that search page. When we researched how to use this tool, we found that there was a Chrome extension for it. It was free for some features and had a limited free trial. Thus, we used this third-party extension to find the sales history of the obsolete accessories. However, we found another problem when we collected the data via AMZ Scout: If we wanted to get the CSV file of the entire product's historical sales data, we needed to subscribe to the monthly or yearly plan, which was expensive. We could only see the sales history chart with the free trial plan. So, we decided to write a script to get the data from the chart. We obtain the data for each point in the line chart by simulating mouse movement and saving it as a CSV file. Then, we could use these data to calculate the quantity and analyze the trend of these obsolete accessories.

Although the method we used was neither the best nor the most accurate, that was the most data we could collect without paying. With this data, we could do a lot of analysis and observe overall product sales trends to achieve the goal we wanted to research. Looking forward to the future, we will have more free tools that can help to get more data, help researchers make more accurate calculations, and analyze past and future trends.

\subsection{Data Analysis}
There are several challenges in the data analysis phase of this project. The main issue is the complexity and variability of sales data across different platforms and time. We face a lack of standardized sales reporting and incomplete data sets since our primary data sources are online retail platforms, particularly Amazon. Although Chrome extensions are used to access Amazon sales history, the granularity and completeness of the data are limited.

Another important issue is the lack of correlation data between sales figures and product release cycles, as well as external factors that influence buying behavior (such as marketing campaigns or economic events). Because of this limitation, it is challenging to directly link sales trends to I/O device obsolescence after the release of the new MacBook Pro.

Furthermore, Apple's official website does not provide sales data, and the company's financial statements do not break out accessory sales for our purposes. Since we do not have enough official sales data, we analyze third-party data instead. However, this may not accurately represent the situation of the real market.

Furthermore, we do not consider the second-hand market. Older devices and accessories may find a home on the secondary market, which could lessen their negative environmental effects. Based on these challenges, we recommend that future research seek more direct sources of data, possibly through partnerships or data-sharing agreements with manufacturers and retailers, in this way to improve the accuracy of sales trend analysis and its correlation with technological obsolescence.

\subsection{Future Work}
As technology develops, people are making faster and faster advances in various areas such as wireless networks, secure communications, smart life \cite{MILLER2022100245,DBLP:journals/corr/abs-2112-15169,YAO2020100087,MILLER2020100089,8556650,10.1145/3127502.3127518,10.1145/3132479.3132480,GAO201718}, machine learning, and so on. In the future, we can combine various current wireless communication technologies \cite{wire1,wire3,10017581, 9523755,9340574,10.1145/3387514.3405861,9141221,9120764,10.1145/3356250.3360046,8737525,8694952,10.1145/3274783.3274846,10.1145/3210240.3210346,8486349,8117550,8057109,https://doi.org/10.1155/2017/5156164,10189210} to improve the accuracy and usability of data by streamlining the process of collecting data from various sources such as product reviews and sales platforms. Advanced secure communication methods \cite{wire2, 10125074,285483,10.1145/3395351.3399367} can ensure the integrity and privacy of sensitive data, especially when working with third-party tools or large organizations such as Amazon or Apple. Additionally, machine learning-based methods \cite{10.1145/3460120.3484766,9709070,9444204,ning2021benchmarkingmachinelearningfast,8832180,8556807,8422243,chandrasekaran2022computervisionbasedparking,iqbal2021machinelearningartificialintelligence,pan2020endogenous} can be used to create predictive models to estimate trends in obsolete devices, optimize environmental strategies, and more accurately assess economic impact.

Since the information is incomplete, it may be helpful for us to subscribe to a third-party tool to get all the data. This would make the analysis more accurate and credible. Due to the expensive payment, we recommend subscribing if you need long-term research. Otherwise, if we have enough to pay with the tools, we can also try other third-party tools that we have mentioned before. Then, we can compare whether different tools will have different results. Then, we can choose the relatively correct one to use for analysis, or we can integrate the results of these different tools. This method may help make the analysis clearer and more credible. Of course, the best way is to cooperate with Amazon or Apple to obtain official data to calculate and analyze.

On the other hand, if we have enough datasets, it may be possible to train a model to calculate sales via product reviews, ratings, prices, etc., in the future. Then, we can use crawler technology to get all the data we need from Amazon and put the data into the model. We will get an estimated product sales. This method will be more convenient for people to get the data. However, it's an estimated number. Thus, it's more suitable for calculating and assessing trends or total product sales.

Additionally, it's possible to train a model to calculate and estimate future trends and total product sales. It will be helpful to evaluate the total number of obsolete I/O devices that will occur because we can only calculate the number from the past to now. We can't know how many obsolete I/O devices will be produced in the future. If there is such a prediction model, it will help us delve deeper into this research. Moreover, we can estimate the quantity of obsolete accessories and the subsequent impact that will occur. It will be a significant study for environmental research. We believe it can help improve the problems of the environment. 

Next, regarding environmental protection strategies, we just studied the research papers and used the theories and methods to analyze our research. We think there are other better ways to develop strategies. For example, create a model that can directly analyze the best strategies for our research problem. It may be very useful because we only read a few research papers, which could only give us a little idea. However, it's not enough for us to develop the best strategy. If we can use a model with the enriched dataset, it may have a great analysis of the problems, provide more ideas, and also select the best method for us. Thus, if this model is developed, we believe it can help many people to complete environmental protection tasks.

Finally, in the economic impact section, we wish that these obsoleted I/O devices not only can be reused but also promote economic growth. Thus, it's helpful to develop a model that can estimate how the obsoleted I/O devices will impact the economy. For instance, with the historical data of these product sales, we can estimate the economic growth caused by these products. Then, estimate when the product becomes obsolete and how it will decrease economic growth. Furthermore, what would be more beneficial to the economy if these discarded products were continued to be sold or recycled directly? Therefore, we need this model to help us analyze the data for the economy. It will be a significant tool for delving into an economic problem for this research question in the future.

\section{Conclusion}
This study investigates the evolution of MacBook Pro I/O devices and their environmental and economic impacts. Through analysis, we can see that while technological progress promotes the advancement of the industry, it also accelerates the elimination of equipment and also brings challenges.

Our research results show that frequent updates of MacBook Pro I/O ports directly lead to shortened life cycles of many accessories. This not only leads to an increase in e-waste but also causes considerable environmental harm due to improper handling and recycling methods. Improved waste management strategies are urgent, as are policies that encourage the development of more durable, universally compatible accessories.

From an economic perspective, the rapid elimination of outdated accessories stimulates consumer spending and industry growth, but it also comes at a cost. It places a financial burden on consumers, who must regularly update their devices. This also leads to the depletion of natural resources. This study highlights the balance that must be achieved between promoting innovation, sustaining the economy, and environmental sustainability.

 We can face these challenges by reuse and recycling strategies. It not only reduce environmental impact, but also provide significant economic benefits. For example, it creates many jobs through recycling and reducing the costs associated with product manufacturing. In addition to improving sustainable economic growth, policies that support product longevity and repairability can reduce the environmental impact of technological advancement.

All in all, the tech industry advances due to the quick technological innovation found in products such as the MacBook Pro, but it also needs a commitment to sustainable practices. Future research should focus on promoting recycling technologies, advocating sustainable product design, and developing effective policies that combine economic growth with environmental protection. Technology development can be made more sustainable and contribute to a more stable economic future for the technology sector by tackling these problems.

\clearpage

\bibliography{mybibfile,zhu}
\end{document}